**Acoustic topological insulator and robust one-way sound transport**

Cheng He[1], Xu Ni[1], Hao Ge[1], Xiao-Chen Sun[1], Yan-Bin Chen[1], Ming-Hui Lu[1,3]†, Xiao-Ping Liu[1,3],

Liang Feng[2], and Yan-Feng Chen[1,3]†

[1]*National Laboratory of Solid State Microstructures & Department of Materials Science and*

*Engineering, Nanjing University, Nanjing 210093, China*

[2.] *Department of Electrical Engineering, University at Buffalo, The State University of New York,*

*Buffalo, NY 14260, USA*

[3]*Collaborative Innovation Center of Advanced Microstructures, Nanjing University, Nanjing, 210093,*

*China*

†Correspondence and request for materials should be addressed to M. H. Lu (luminghui@nju.edu.cn)

and Y. F. Chen (yfchen@nju.edu.cn).

**Discovery of novel topological orders of condensed matters is of a significant interest in both fundamental and applied physics due to the associated quantum conductance behaviors and unique symmetry-protected backscattering-immune propagation against defects, which inspired similar fantastic effects in classical waves system, leading to the revolution of the manipulation of wave propagation. To date, however, only few theoretical models were proposed to realize acoustic topological states. Here, we theoretically and experimentally demonstrate a two dimensional acoustic topological insulators with acoustic analogue of quantum spin Hall Effect. Due to the band inversion mechanism near the double Dirac cones, acoustic one-way pseudospin dependent**

**propagating edge states, corresponding to spin-plus and spin-minus, can be observed at the interface between two graphene-like acoustic crystals. We have also experimentally verified the associated topological immunity of such one-way edge states against the different lattice defects and disorders, which can always lead to inherent propagation loss and noise. We show that this unique acoustic topological phenomenon can offer a new promising application platform for the design of novel acoustic devices, such as one-way sound isolators, acoustic mode switchers, splitters, filters etc.**

The concept of topology is originated from mathematics to characterize the geometrical property of objects. In physics, topological description of electronic quantum behaviors can set a new paradigm in the classification of condensed matters. For example, ordinary insulators have complete band gap with trivial topological order. But in the presence of magnetic field with broken time-reversal symmetry (TRS), quantum Hall effect (QHE), possessing gapless edge state conducting upper and lower bulk bands resulting in one-way charge transport at the boundary, can be characterized by topological Chern number (1, 2). Without breaking TRS, another class of materials, topological insulator (TI) or saying quantum spin Hall effect (QSHE) can be obtained with spin-orbit coupling, supporting spin transport at the boundary, that is opposite spins one-way propagating in opposite direction against non-magnetic defects, which can be characterized by topological Z2 order or spin Chern number (3-6). The hallmark of these topological states is their quantum behaviors and unique robust one-way propagation, which can be of great importance not only in fundamental science but also in application and engineering.

Since the design concept of topological engineering (TRS broken or conserved) is generic, topological states can also be extended to other types of classical waves beyond electronic regime,, such as electromagnetic wave (7-17), mechanical vibrations (18, 19), elastic wave (20, 21) and acoustic wave

(22-26). In photonics, recently experimentally realized photonic QHE (9, 10) and Floquet QHE (13) are based on broken TRS of magneto-optical photonic crystals or broken z-direction symmetry of helical structure. By using two degenerate polarization states (14, 16, 27, 28), photonic modes (17, 29), or helicity of energy flows as pseudospins (11, 12, 17), photonic QSHE and TI have also been proposed and partially realized (12, 16). However, for longitudinal acoustic wave, e.g. sound, to break TRS or construct degenerate modes is more difficult than that in photonic system due to weak magneto-acoustic effects and only one polarization state of longitudinal sound. To date, only few theoretical models have been proposed to realize acoustic topological states, using circulating fluid (22-25), gyroscope (20) and magneto-elastic effects with broken TRS to mimic QHE, or using phononic ring resonators (crystal) consisted of metamaterials constituents as analogue of QSHE (21, 26). It is more desirable to design external "force"-free acoustic topological states with simple structure, leading to experimental realization of robust one-way sound transport.

Here, we focus on designing an experimentally feasible acoustic crystals (ACs) to realize topological states and robust one-way sound transport without external "force". As shown in Fig. 1a, a sound waveguide can be formed at the interface between two graphene-like (honeycomb-lattice) ACs consisted of stainless steel in air background, with the same lattice constant but different radii of rods ($r$ is radius of rod and $a$ is center to center distance of the nearest neighbor rods). The proposed waveguide has nontrivial topological property, which means that the sound will propagate in this waveguide with 100% transmission and no backscattering, robust against various kinds of defects, i.e. cavity, disorder and even sharp bends.

On the other hand, the key point to realize classical wave (spin-1) analogue of QSHE in electronics (spin-1/2) with robust transport is to increase degrees of freedom to two-fold states which experience two opposite gauge fields to mimic spins and spin-orbit couplings. Thus, it is necessary to construct four-fold degenerate states for acoustic states to mimic the Dirac cone of electrons where the QSHE occurs around. Fortunately, the coincident degeneracy of two Dirac cones, that is double Dirac cones, can be realized at the Brillouin (BZ) center of graphene-like (or triangle-lattice) acoustic (or photonic) crystal system (*26-28*). According to group theory, such lattice possesses $C_{6v}$ symmetry having two two-dimensional irreducible representations which could meet together with judiciously designed parameters. At the BZ center, there exists four-fold degenerate acoustic states like the Dirac point in electronic system, which can be treated as a good candidate to construct acoustic QSHE. Furthermore, these four-fold degenerate acoustic states will split into two two-fold degenerate states and open a bulk gap by changing the structure parameters. Fig. 1b is a schematic of band structures and acoustic states with decreasing the radii of rods (filling ratio) in graphene-like ACs. When *r/a*=0.45 (the left panel of Fig. 1b), two two-fold degenerate acoustic states of lower and upper bulk bands appear at the BZ center, corresponding to $p_x/p_y$ and $d_{x^2-y^2}/d_{xy}$, respectively. Similar to *p* and *d* orbitals of electrons, $p_x$ obeys symmetry $\sigma_x/\sigma_y$=-1/+1; $p_y$ obeys $\sigma_x/\sigma_y$ =+1/-1; $d_{x^2-y^2}$ obeys $\sigma_x/\sigma_y$ =+1/+1; $d_{xy}$ obeys $\sigma_x/\sigma_y$ =-1/-1, where $\sigma_x/\sigma_y$ =+1/-1 means the even/odd parity of spatial inversion relative to *x/y* axis. When *r/a* decreased to 0.3928 (the middle panel of Fig. 1b), double Dirac cone is formed with four-fold degenerate states, where lower and upper bands touch at one point with absence of bulk band gap. Further decreasing the filling ratio will reopen a bulk band gap as shown in the right panel of Fig. 1b, *r/a* =0.3 for instance. Surprisingly, the two-fold degenerate states of upper bands change to $p_x$ and $p_y$, while other two of lower bands are $d_{x^2-y^2}$ and $d_{xy}$. Such phenomenon indicates that band conversion mechanism takes place with decreasing the radii of rods. Hybridizing four states mentioned above as $p_{\pm} = (p_x \pm ip_y)/\sqrt{2}$ and $d_{\pm} = (d_{x2-y2} \pm id_{xy})/\sqrt{2}$ can create pseudospins for acoustic wave (*29*). Thus, the topological transition can take place at

$r/a$=0.3928, that is the point of double Dirac cones, from ordinary acoustic crystal (OAC) to topological acoustic crystal (TAC) upon adjusting the filling ratio (details see Figs. S1 and S2 in supplementary).

Then, we calculate the projected band structure ($a$=1cm) with a configuration of TAC ($r/a$=0.3) adjacent to OAC ($r/a$=0.45) shown in Fig. 1c. The density and longitudinal sound speed of stainless steel (air) are 7800 $kg/m^3$ (1.25 $kg/m^3$) and 6010 $m/s$ (343 $m/s$). A pair of topological edge states (red and blue lines) can be observed in the overlapped bulk band gap (shadow regions) of these two ACs, frequency ranging from 19.004 kHz to 20.528 kHz with 7.7% relative band width, supporting sound propagation at the interface of two ACs. Two acoustic spins are hybridized by the mode of symmetrical (S) real part and anti-symmetrical (A) imaginary part (right panel), corresponding to acoustic spin+ $S+iA$ and spin- $S-iA$, respectively. Such hybridizations provide pseudo-spin for longitudinal acoustic wave, which can give rise to acoustic spin and spin-orbit coupling. And, each individual acoustic spin edge state has the same sign of group velocities (slope direction) in the whole BZ, indicating the existence of one-way propagation. Meanwhile, acoustic spin+ (positive slope) and spin- (negative slope) (*29*) with opposite signs means that opposite spins one-way propagate in opposite direction along the boundary. Therefore, acoustic spin+/- only can clockwise/anti-clockwise propagate around the TAC boundary in one-way, which is an acoustic-wave counterpart of QSHE with spin-dependent transport.

To verify the one-way spin-dependent transport of acoustic spins in experiment, we design a cross-waveguide splitter (*30*) as shown in Fig, 2a. The dashed lines divide the whole area into four parts with four edges used as input and output ports (labeled as 1, 2, 3, 4): the upper and lower areas are TAC ($r/a$ =0.3), the left and right areas are OAC ($r/a$ =0.45). According to spin-dependent one-way clockwise/anti-clockwise circulating effect analyzed above, ports 1 and 3 can only support spin- input and spin+ output while ports 2 and 4 can only support spin- output and spin+ input. It is attributed to one-way clockwise

(blue circular arrow in Fig. 2a) and anti-clockwise (red circular arrow in Fig. 2a) propagation for acoustic spin+ and spin- around the TAC boundary. If we use port 1 as input port for acoustic spin-, the sound can only come out through ports 2 and 4, and there will be no sound detected at port 3. As a contrastive case, when using acoustic spin+ incidence from port 2, sound cannot be detected at port 4, and vice versa for other ports. Fig. 2b and Fig. 2c shows the simulated pressure filed distribution of these two cases at 19.8 kHz consistent with our discussion. Corresponding experimental transmission spectra are shown in Fig. 2d and Fig. 2e. Pij represent the transmission at port j incidence from port i (i=1,2 and j=1,2,3,4). We can find the opposite spins are extremely suppressed (red line in Fig. 2d and blue line in Fig. 2e), up to 30 dB compared to the transmission of supporting spins (blue line in Fig. 2d and red line in Fig. 2e). The small sound transmission out of forbidden ports (P13 for spin- and P24 for spin+) is due to the limited width of the bulk band gap, which can be further optimized by using more rods in experiments. Thus, the one-way spin-dependent transport of acoustic QSHE is well demonstrated in this cross-waveguide splitter model, which can be used in acoustic one-way splitter, switch and acoustic spin filter devices.

One of the most striking features of topological states is their robust one-way propagation against non-magnetic defects (non-spin-mixing defects in our model), which has huge potential applications in wave guiding, sound communication beyond traditional ways. Therefore, it is necessary to check the robustness of such acoustic topological model against defect. As a comparison, we also construct an ordinary waveguide by removing a row of rods in OAC. Fig, 3a is a picture of our sample, where the green dashed line shows the topological waveguide constructed by the interface between TAC ($r/a$ =0.45) and OAC ($r/a$ =0.3) while the yellow dashed line shows the ordinary waveguide in OAC. Three kinds of defects, small cavity (4 rods removed), disorder and bends, are used to check the robustness. Fig. 3b is the simulated pressure distribution with such three kinds of defects at 20.1 kHz. In topological cases, using acoustic spin- incidence from left, there are no obvious resonances when sound transmit through these defects, and the transmission is kept the same. But in ordinary cases, resonances can be excited due to the defect, largely decreasing the transmission even with total reflection of sound. Fig. 3c is the experimentally measured transmission spectra in topological waveguide compared to that in ordinary waveguide as shown in Fig. 3d. Shadow regions represent the bulk band frequency area. Black, red, blue and green lines represent the transmission spectra without defects, with cavity, disorder and bends, respectively. It can be clearly observed that the transmission almost has no change in topological case in the edge state region while the transmission is largely decreased with several resonant dips in ordinary-waveguide cases. The average transmission of topological case is around -5 dB. A little difference between the transmission spectrum with defects and without defect in topological case may come from the intrinsic loss: the longer distance the sound will experience around the defects.

To conclude, we provide a convenient way to construct the acoustic topological states with one-way robust spin-dependent transport based on band conversion mechanism, which can be treated as an acoustic QSHE. Such phenomena obtained in classical system may help to understanding quantum behavior of classical waves and pave a new way to studying topological properties in ACs. The advantage of our model includes the external "force"-free setup and easy fabrication. The effect we obtained can be largely scalable from audible sound to ultrasonic frequencies. Our realized prototype can be directly used to realize the similar phenomena for underwater sound by using water background. We believe that this unique topological phenomenon of sound can immediately offer promising applications such as sound guiding, communication, switch, splitter and acoustic spin filter.

**Figure captions**

**Figure 1 | Schematic of acoustic topological states and band conversion mechanism. a,** Robust propagation of sound against defects in topological waveguide constructed by two graphene-like ACs consisted of stainless steel in air background. The inset shows zoom-in picture, where *r* is radius of rod and *a* is center to center distance of the nearest neighbor rods. **b,** Band conversion mechanism: two two-fold degenerate acoustic states (*r/a*=0.45) with bulk gap; to four-fold degenerate (*r/a*=0.3928) with absence of gap; then reopen gap (*r/a*=0.3) when continuously decreasing the filling ratio. The topological transition from OAC to TAC occurs near double Dirac cones. $p_x$, $p_y$, $d_{x^2-y^2}$, $d_{xy}$, represent different parities of acoustic states. Band structures are shown in black curves. **c,** Projected band structure (*a*=1cm) with TAC adjacent to OAC. Red and blue lines represent the acoustic spin+ and spin- hybridized by S and A modes (right panel show modes at $k_∥$=0.05). Shadow regions represent the bulk band frequency area.

**Figure 2 | Acoustic one-way spin-dependent transport. a,** A picture of sample used in experiment: the upper and lower areas are TAC, the left and right areas are OAC. Four ports are labeled as 1, 2, 3, 4. Only clockwise circulating propagation is allowed for acoustic spin+ (red circular arrow), and vice versa for spin- (blue circular arrow). Simulated pressure filed distribution at 19.8 kHz with acoustic wave incidence from **b,** port 1 and **c,** port 2. Experimental transmission spectra with **d,** acoustic spin- incidence from port 1 and **e,** acoustic spin+ incidence from port 2. Pij represent the transmission at port j incidence from port i (i=1,2 and j=1,2,3,4). Shadow regions represent the bulk band frequency area.

**Figure 3 | Robust one-way sound transport. a**, A picture of experimental setup. Green (yellow) dashed lines represent topological (ordinary) waveguides. **b,** Simulated pressure distribution against three types of defects: cavity, disorder and bends using acoustic spin- incidence at 20.1 kHz. Blue, red and black arrows represent propagation direction of acoustic spin-, spin+ and ordinary sound, respectively.

Experimental transmission spectra **c,** in topological waveguide compared to that **d,** in ordinary one. Black, red, blue and green lines represent the transmission spectra without defects, with cavity, disorder and bends, respectively. Shadow regions represent the bulk band frequency area.

Fig. 1

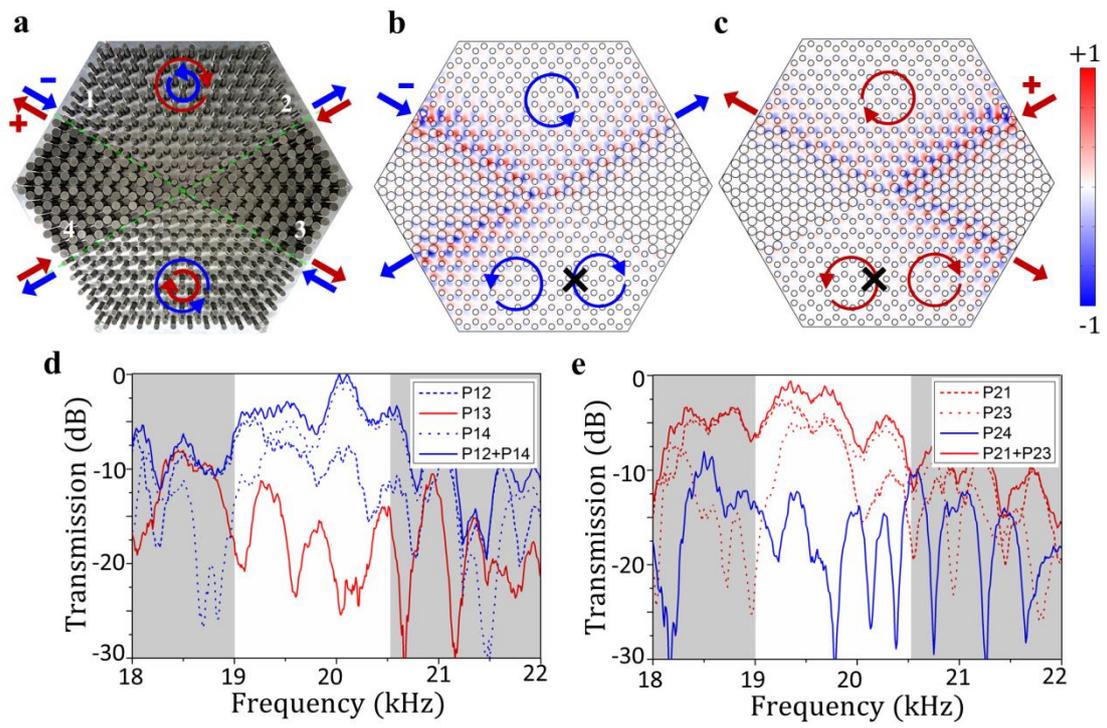

Fig. 2

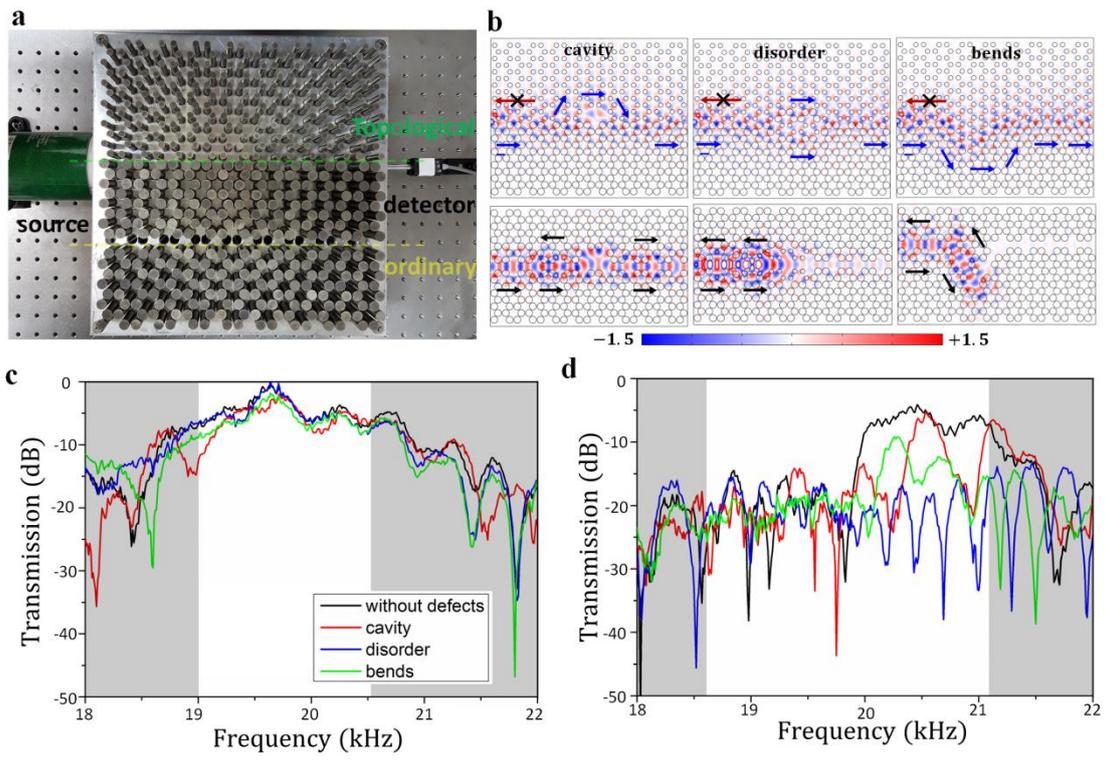

Fig. 3